\documentclass{article}
\usepackage{theorem, amssymb, amsmath}

{\theorembodyfont{\rmfamily}
       \newtheorem{definition}{Definition}
       \newtheorem{defn}[definition]{Definition}

    \newtheorem{text-conjecture}[definition]{Conjecture}

\theoremstyle{plain}
    
    \newtheorem{lemma}[definition]{Lemma}
    \newtheorem{theorem}[definition]{Theorem}

\newcommand{\proofof}[1]{{\bf Proof of #1:} }
\newcommand{\proofend}{\hfill$\blacksquare$}

\newcommand{\calA}{{\mathcal A}}
\newcommand{\calF}{{\mathcal F}}
\newcommand{\ceil}[1]{{\left\lceil #1 \right\rceil}}
\newcommand{\floor}[1]{{\left\lfloor #1 \right\rfloor}}
\newcommand{\E}{{\mathbf E}}
\newcommand{\R}{{\mathbb R}}
\newcommand{\qu}[1]{\left| #1 \right\rangle}

\title{Quantum lower bound for the collision problem}
\author{Samuel Kutin\thanks{Center for Communications Research,
805 Bunn Drive, Princeton, NJ 08540.
Email: {\tt kutin@idaccr.org}.}}

\begin{document}
\maketitle
\begin{abstract}
We extend Shi's 2002 quantum lower bound for collision in $r$-to-one
functions with $n$ inputs.
Shi's bound of $\Omega((n/r)^{1/3})$ is tight, but his proof
applies only in the case where the range has size at least
$3n/2$.  We give a modified version of Shi's
argument which removes this restriction.
\end{abstract}

\section{Introduction}
How many quantum queries does it take to find a collision?  Many
cryptographic systems depend on the difficulty of finding collisions,
so it is important to understand how difficult this problem may prove
for a quantum computer.

Obviously, it may be easier to find collisions in some functions
then others.  We are interested in a {\em black-box\/} argument:  our
only access to the function is as a quantum oracle.  We are promised
that the function is $r$-to-one.  (We require that $r$ be a divisor of
$n$, the size of the input space.)  Brassard, H{\o}yer, and
Tapp \cite{BHT97} gave a quantum algorithm which requires
$O((n/r)^{1/3})$ quantum queries, an improvement over the
$\Theta((n/r)^{1/2})$ classical queries needed.  In this note,
we are concerned with the matching lower bound.

For a lower bound, it is easier to consider a decision problem:
the input function is guaranteed to be either one-to-one or
$r$-to-one, and our task is to determine which case holds.
Aaronson \cite{Aar02} proved the first significant lower bound:
$\Omega((n/r)^{1/5})$ queries.

More recently, Shi \cite{Shi01} proved a lower bound of
$\Omega((n/r)^{1/3})$, given the additional condition that the size
of the range of the function is at least $3n/2$.  (In the case where
the range is only $n$, Shi provides a lower bound of $\Omega((n/r)^{1/4})$).
Shi's proof is a novel application of the methods of
Nisan and Szegedy \cite{NS92} to the case where one cannot fully
symmetrize the multivariate polynomials.

Our main result is a new version of Shi's theorem, but without the
additional constraint on the size of the range:

\begin{theorem}
\label{main-thm}
Let $n > 0$ and $r \ge 2$ be integers with $r \mid n$, and let a function
from $[n]$ to $[n]$ be given as an oracle with the promise that it is
either one-to-one or $r$-to-one.  Then any quantum algorithm for
distinguishing these two cases must evaluate the function
$\Omega\left((n/r)^{1/3}\right)$ times.
\end{theorem}

The argument is very similar to that of Shi.  As stated above, we
remove the requirement that the range be at least $3n/2$.  Our
proof is conceptually simpler for other reasons:
\begin{enumerate}
\item The natural automorphism group on the set of functions
from $[n]$ to $[N]$ is $S_n \times S_N$.  Our argument symmetrizes with
respect to the entire group.
\item We avoid the explicit introduction of the problem Half-$r$-to-one.
\end{enumerate}

\section{Preliminaries}

\subsection{Functions as quantum oracles.}

Let $n, N > 0$ be integers.  Let $\calF(n,N)$ be the set of functions
from $[n]$ to $[N]$.

Our functions are given to us as a quantum oracle.  We can perform
a transformation $O_f$, which applies $f$ to the contents of some of
the quantum state:
$$
O_f \qu{i,j,z} = \qu{i, f(i) + j \pmod N, z}.
$$
Here $z$ is a placeholder for the unaffected portion of the quantum state.

The query complexity of a quantum algorithm is the number of times it
calls $O_f$.  We think of our algorithm as alternating between $T+1$
unitary operators and $T$ applications of $O_f$.

Let $\delta_{i,j}(f)$ be $1$ when $f(i) = j$.  Then, after $T$ queries,
the amplitude of each quantum base state is a degree-$T$ polynomial
in these $\delta_{i,j}(f)$.  Hence, the acceptance probability $P(f)$
is a polynomial over $\delta_{i,j}$ of degree at most $2T$.
This connection between quantum complexity and polynomial degree
is due to Beals, et al.\ \cite{BBCMdW}.

Note that this polynomial $P(f)$ is constrained to be in the interval
$[0,1]$ whenever the $\delta_{i,j}$ correspond to a valid input; i.e.,
\begin{align}
\notag
\forall i, j, \qquad & \delta_{i,j} \in \{0, 1\} \\
\label{delta-constraint} \forall i, \qquad & \sum_j \delta_{i,j} = 1
\end{align}

The connection between polynomial degree and query complexity
was first made by Nisan and Szegedy \cite{NS92}.  In their applications,
they symmetrized over all permutations of the variables, reducing the
multivariate polynomial to a univariate polynomial.  They then apply
results from approximation theory to prove a lower bound on the
degree of the polynomial.  Beals, et al.\ \cite{BBCMdW} followed the
same approach.

A nice, general version of the approximation theory results was
shown by Paturi \cite{Pat92}.  Following Shi \cite{Shi01}, we use
a slight modification of Paturi's theorem:

\begin{theorem}[Paturi]
\label{Paturi-thm}
Let $q(\alpha) \in \R[\alpha]$ be a polynomial of degree $d$.
Let $a$ and $b$ be integers, $a < b$, and let $\xi \in [a,b]$ be a
real number.
If
\begin{enumerate}
\item $|q(i)| \le c_1$ for all integers $i \in [a,b]$, and
\item $|q(\floor{\xi}) - q(\xi)| \ge c_2$ for some constant $c > 0$,
\end{enumerate}
then
$$
d = \Omega(\sqrt{(\xi - a + 1)(b - \xi + 1)}),
$$
where the hidden constant depends on $c_1$ and $c_2$.
\end{theorem}

Note that, if the conditions of the theorem are met for any $\xi$,
we have $d = \Omega(\sqrt{b - a})$.  If they are met for some
$\xi \approx (a+b)/2$, then $d = \Omega(b - a)$.

In our setting, the automorphism group for the variables $\delta_{i,j}$
is $S_n \times S_N$.  If we symmetrize with respect to this group, we
do not immediately obtain a univariate polynomial.  Hence, we will have
to work harder to apply Theorem~\ref{Paturi-thm}.

For $\sigma \in S_n$, $\tau \in S_N$, we define
$\Gamma^\sigma_\tau \colon \calF(n,N) \to \calF(n,N)$ by
$$
\Gamma^\sigma_\tau(f) = \tau \circ f \circ \sigma.
$$

Let $P(f)$ be an acceptance polynomial as above.  We can write
$P$ as a sum $\sum_S C_S I_S(f)$, where $S$ ranges over subsets
of $[n] \times [N]$, and
$$
I_S = \prod_{(i,j) \in I_S} \delta_{i,j}.
$$
By (\ref{delta-constraint}), we may assume that each pair
$(i,j) \in S$ has a distinct value of $i$; we thus write
\begin{equation}
\label{monomial-form}
I_S = \prod_{k=1}^t \prod_{i \in S_k} \delta_{i,j_k},
\end{equation}
where the sets $S_k$ are disjoint, and $\sum_k |S_k|$ is the degree
of the monomial.

\subsection{Some special functions}

We now define a collection of functions which are $a$-to-one on
part of the domain, and $b$-to-one on the rest of the domain.
(These will enable us to interpolate between one-to-one and
$r$-to-one functions.)

Fix $N \ge n > 0$.  We say that a triple $(m,a,b)$ of integers
is {\em valid\/} if $0 \le m \le n$,
$a \mid m$, and $b \mid (n - m)$.  For any such valid triple, we
have a function $f_{m,a,b} \in \calF(n,N)$, given by
$$
f_{m,a,b} = \begin{cases}
\ceil{i / a} & 1 \le i \le m, \\
N - \floor{(n - i)/b} & m < i \le n.
\end{cases}
$$
So $f_{m,a,b}$ is $a$-to-one on $m$ points, and
$b$-to-one on the remaining $n - m$ points.  (Since $N \ge n$,
the two parts of the range do not overlap.)

Note that our $f_{m,a,b}$ plays the same role as Shi's
$f_{m,g}$, with $a = g$ and $b = 2$.

We now examine the behavior of $f_{m,a,b}$ after we symmetrize by
all of $S_n \times S_N$.

\begin{lemma}
\label{Q-lemma}
Let $P(f)$ be a degree-$d$ polynomial in $\delta_{i,j}$.
For a valid triple $(m,a,b)$, define $Q(m,a,b)$ by
$$
Q(m,a,b) = \E_{\sigma, \tau}
\left[ P\left(\Gamma^\sigma_\tau(f_{m,a,b})\right)\right].
$$
Then $Q$ is a degree-$d$ polynomial in $m,a,b$.
\end{lemma}

\begin{defn}
For integers $k, \ell$, let $\ell^{\bar k}$ denote the falling
power $\ell(\ell - 1) \cdots (\ell - k + 1)$.
\end{defn}

\proofof{Lemma~\ref{Q-lemma}}
It suffices to prove the lemma in the case where $P$ is a monomial
$I_S$.  We write $I_S$ in the form~(\ref{monomial-form}); then
$d = |S|$.  We write $s_k = |S_k|$.

For each subset $U \subseteq [t]$, let $A_U$ be the following
event:  for each $k \in U$, $\sigma^{-1}(j_k) \le m/a$;
for each $k \notin U$, $\sigma^{-1}(j_k) \ge N - (n-m)/b + 1$.

Clearly the events $A_U$ are disjoint.  If $I_S\left(\Gamma^\sigma_\tau
(f_{m,a,b})\right)$ is nonzero, then every $\sigma^{-1}(j_k)$ must lie
in the range of $f_{m,a,b}$, so some event $A_U$ must occur.  Hence,
we write

\begin{align*}
Q(m,a,b) & = \sum_{U \subseteq [t]} \Pr(A_U) Q_U(m,a,b),
\intertext{where}
Q_U(m,a,b) & = \E_{\sigma, \tau}\left[
I_S\left(\Gamma^\sigma_\tau(f_{m,a,b})\right) \mid A_U \right].
\end{align*}

Choose some $U$, and let $u = |U|$.  Then $\Pr(A_U)$ is given
by
$$
\Pr(A_U) = {\left(m \over a\right)^{\overline u}
\left(n - m \over b\right)^{\overline{t - u}} \over N^{\overline t}},
$$
which is a rational function in $m, a, b$.  The numerator has
degree $t$, and the denominator is $a^u b^{t-u}$.

Also,
$$
Q_U(m,a,b) = {1 \over n^{\overline s}}
\prod_{k \in U} a^{\overline {s_k}} \prod_{k \notin U} b^{\overline {s_k}}.
$$
This is a polynomial in $a,b$ of degree $d$; furthermore
$Q_U$ is divisible by $a^u b^{t-u}$.

Hence, for each $U$, $\Pr(A_U) Q_U$ is a degree-$d$ polynomial in
$m,a,b$.  Therefore $Q(m,a,b)$ is itself a degree-$d$ polynomial.
This concludes the lemma.
\proofend

\section{Main Proof}

We are now ready to prove Theorem~\ref{main-thm}.

\proofof{Theorem~\ref{main-thm}}
Let $\calA$ be an algorithm which distinguishes one-to-one from
$r$-to-one in $T$ queries, and let $P(f)$ be the corresponding
acceptance probability.  $P(f)$ is a polynomial in $\delta_{i,j}$
of degree at most $2T$.  Let $Q(m,a,b)$ be formed from $P$ as in
Lemma~\ref{Q-lemma}, and let $d = \deg Q$; we have $d \le 2T$.

For any $\sigma, \tau$, we know that $\Gamma^\sigma_\tau(f_{m,a,b})$
is a valid function.  If $a = b$, this function is $a$-to-one.
We conclude the following:

\begin{enumerate}
\item $0 \le Q(m,a,b) \le 1$ whenever $(m, a, b)$ is a valid triple.
\item $0 \le Q(m,1,1) \le 1/3$ for any $m$.
\item $2/3 \le Q(m,r,r) \le 1$ for any $m$ where $r \mid m$.
\end{enumerate}

The remainder of the proof consists of proving that
$\deg Q = \Omega(n/r)^{1/3}$.
For simplicity of exposition, we begin with the case $r = 2$.

Let $M = 2\floor{n/4}$.  We ask:  is $Q(M,1,2) \ge 1/2$?
In other words:  does our algorithm accept (at least half the time)
an input which is one-to-one on half the domain, and
two-to-one on the other half?

Case I:  $Q(M,1,2) \ge 1/2$.  Let $c$ be the least integer for
which $|Q(M,1,c)| \ge 2$.  Then we have $Q(M,1,x)$ between $-2$ and $2$
for all positive integers $x < c$, and $|Q(M,1,1) - Q(M,1,2)| \ge 1/6$.  By
Theorem~\ref{Paturi-thm}, we have $d = \Omega(\sqrt{c})$.

Now, we consider the polynomial $h(i) = Q(ci, 1, c)$.  For any integer
$i$ in the range $0 \le i \le \floor{n/c}$, we have $0 \le h(i) \le 1$.
But $|h(M/c)| \ge 2$.  We conclude, by Theorem~\ref{Paturi-thm}, that
$d = \Omega(n/c)$.

Case II:  $Q(M, 1, 2) < 1/2$.  Now, let $c$ be the least even integer
for which $|Q(M, c, 2)| \ge 2$.  As in Case I, we first get
$d = \Omega(\sqrt{c})$.  Then, by considering $h(i) = Q(ci, c, 2)$,
we obtain $d = \Omega(n/c)$.

In either case, by combining $d = \Omega(\sqrt{c})$ and
$d = \Omega(n/c)$, we get $d^3 = \Omega(n)$, or $d = \Omega(n^{1/3})$.

For general $r$, the setup is almost identical:  we now split into
cases based on whether $Q(m,1,r) \ge 1/2$?  (Note that, in Case II, we let $c$
be the least multiple of $r$ for which $Q(M, c, r) \ge 2$.)
We first get
$d = \Omega(\sqrt{c/r})$, and then $d = \Omega(n/c)$, yielding
$d = \Omega((n/r)^{1/3})$.
\proofend

\bibliography{qlower}
\bibliographystyle{plain}

\end{document}